\documentclass[10pt,a4paper]{article}

\title{
Quantum Monte Carlo study of the H$^-$ impurity in small helium clusters
}

\author{Mose Casalegno $^a$, Massimo Mella $^b$ , and Gabriele Morosi $^c$ \ \\
Dipartimento di Chimica Fisica ed Elettrochimica,\\ Universita' degli Studi
di Milano, via Golgi 19, 20133 Milano, Italy\\
$^a$ Electronic address: Mose.Casalegno@unimi.it\\
$^b$ Electronic address: Massimo.Mella@unimi.it\\
$^c$ Electronic address: Gabriele.Morosi@unimi.it \\
\and 
Dario Bressanini\\
Dipartimento di Scienze Chimiche, Fisiche e Matematiche,\\
Universita' dell'Insubria, polo di Como, \\
via Lucini 3, 22100 Como, Italy\\
Electronic address: dario@fis.unico.it\\}

\begin{document}

\maketitle

\bigbreak
\begin{abstract}
We report ground state energies and structural properties for small helium
clusters ($^4$He) containing an H$^-$ impurity computed by means of
variational and diffusion Monte Carlo methods.
Except for $^4$He$_2$H$^-$ that has a noticeable contribution from collinear
geometries where the H$^-$ impurity lies between the two $^4$He atoms,
our results show that our $^4$He$_N$H$^-$ clusters have a compact 
$^4$He$_N$ subsystem that binds the H$^-$ impurity on its surface.
The results for $N\geq 3$ can be interpreted invoking
the different features of the 
minima of the He-He and He-H$^-$ interaction potentials.

\end{abstract}

\pagebreak

\section{INTRODUCTION}

Weakly bound atomic and molecular clusters represent an interesting and growing
field of research in both chemistry and physics ~\cite{whrev}.
They are useful to understand the evolution of the properties
from microscopic systems to bulk matter.
Moreover, they generate alluring questions whose answers are not
trivial due to the important interplay between dynamical
and geometrical factors in the cluster description ~\cite{gianhe3}.
Among the most studied systems, rare gas clusters possess a rich and 
intriguing set of properties directly related to the weakness of the
interaction between their costituent atoms ~\cite{whrev}. 
The shallow well of their interaction potential energy surface (PES) 
allows the rare gas atoms in a cluster to
have large amplitude vibrational motions, therefore sampling all the
features of the PES itself, and precluding the use of the usual
harmonic approximation to describe their vibrational energy levels
~\cite{night}.

Moreover,
the clusters of the lightest rare gases undergo a solid-liquid 
transition at temperatures
of the order of few kelvins, so they represent good candidates as a medium
where reactions can take place, allowing to study and develop low temperature
chemistry. Unfortunately, a direct spectroscopic study of the droplets of
rare gases to acquire accurate information about their internal dynamics
is not easy to carry out, due to the absence of any chromophoric
unit ~\cite{whrev}. Furthermore, molecular beam experiments generate
a to low concentration of small clusters to allow neutron scattering studies
of the internal structure.

Recently, after the discovery that rare gas clusters can easily 
pick up one or more atomic and molecular impurities ~\cite{ton1}, attention
has been paid to study the effect of the impurity on the cluster
and {\it vice versa} ~\cite{whh2,whsf6,chinsf6,lewhf,tonsf6}. 
These studies focused especially on the 
spectroscopic properties of the impurity in the cluster medium,
as a tool to probe the dynamics of the cluster itself.
The necessity to supplement the experimental spectroscopic results
with an interpretation of the measured properties has recently
renewed the theoretical interest on these species. The test of
the accuracy of the available PES describing the interaction between
the various atoms and molecules, and the calculation of the 
effect on the measured spectroscopic quantities
of the increase of the dimension of the cluster 
( i.e. the number of rare gas atoms) 
are the most frequent computational studies ~\cite{lewhf,lewarhf}. 

Among the rare gas atoms, $^4$He owes its importance to the 
strong quantum features it displays in clusters and in liquid bulk
at low temperatures ~\cite{whrev}.
These features are responsible for the macroscopic superfluid behavior
of $^4$He, that manifests itself in the total absence of viscosity and
in the ability to quickly transport the heat from a source to the surronding
matter~\cite{gordon}.
A recent experiment on the OCS molecule absorbed in small
$^4$He clusters has definitively shown that superfluidity can be 
found even in microscopic aggregates, putting to an end a long debate
~\cite{tonocs}.
Moreover, the low temperature of these clusters can help spectroscopic
studies on large molecules: the electronic
spectra of the aminoacids tryptophan and  tyrosine were simplified by
cooling their vibrational motion inside an $^4$He droplet ~\cite{tonacid},
llowing an easier interpretation of the experimental results.
While alkali and alkali-earth atoms are adsorbed
on $^4$He clusters and then investigated by electronic spectra
~\cite{sticasr,stilinak}, all other impurities, studied by means of
infra-red spectroscopy, reside inside the clusters themselves
~\cite{whsf6,tonocs}, and strongly
perturbate their structure and properties. This difference is due to the
weaker or
stronger interactions between the helium atom and the doping molecule than
between two helium atoms.
Different information on the total dynamics of these aggregates
could be extracted if the impurity pertubates only slightly the 
cluster, or even if, when attached to the droplet, it generates a system
whose global properties can be studied by means of microwave or infra-red
techniques.

Recently, a high accuracy PES for the helium-hydride ion interaction has
become available ~\cite{bend}: the main features of this interaction potential
are the small depth (about 4 cm$^{-1}$), and the large value 
of the distance where the
minimum is located (about 13 bohr). These features show themselves in the
wide amplitude motion of the dimer ( the mean value for the $^4$He-H$^-$ 
distance
is about 22 bohr), and in the long tail of the vibrational ground state
wave function. An interesting property of this quantum system is given by
the existence of an excited rotational state with J=1 ~\cite{bend}. 
The weak interaction between He and H$^-$ makes the hydride anion a good
candidate for studies on the He clusters weakly perturbed by an impurity.
Due to the presence
of a negative charge that is able to polarize the helium atom, 
this system should
have a finite dipole moment, hence to be microwave active.
Moreover, because of the small mass difference between $^4$He 
and H$^-$,
the spectroscopic techniques should probe the quantum motion of 
the whole system. This could
be the case also for clusters containing more helium atoms, allowing,
as stated before, to collect different information about the
quantum dynamics of these aggregates.

In this work we studied both the energetics and structures of the ground state
of the $^4$He$_N$H$^-$ clusters using
quantum Monte Carlo (QMC) methods. 
During the last few years, these methods have been proved
to give quite accurate information
even for highly quantum systems like helium clusters 
~\cite{whcluster,rick,lewhe}.
The main goal of this study is to obtain a clear picture of the relative motion 
and distribution between the $^4$He and the H$^-$ species,
with a special emphasis on the location of the H$^-$ impurity with respect
to the $^4$He$_N$ moiety.

The remainder of the paper is organized as follows. Section II contains
the description of the theoretical approach, of the interaction
potentials used and few comments about the Monte Carlo simulations.
Section III contains the discussion of the Monte Carlo results, while
section IV reports our conclusions and possible future directions
of this study.


\section{METHODS}

In atomic units,
the Hamiltonian operator 
for any $^4$He$_N$H$^-$ clusters containing N $^4$He atoms and one H$^-$ ion
is

\begin{equation}
\label{cluseq1}
{\mathcal H} = -\frac{1}{2} \left( \sum _{i=1}^{N}
\frac{\nabla _i ^{2}}{m_{^4 He}}
+ \frac{\nabla_{H^-}^{2}}{m_{H^-}} \right)
+ V(\mathbf{R})
\end{equation}

\noindent
where $V(\mathbf{R})$ is the interaction potential and
$\mathbf{R}$ is a point in configuration space. Here, we
assume a pair potential of the form

\begin{equation}
\label{cluseq2}
V(\mathbf{R}) = \sum _{i<j} V_{HeHe}(r_{ij})
+ \sum _i V_{HeH^-}(r_i)
\end{equation}

\noindent
where
$r_{ij}$ is the distance between the i-th and j-th helium atom,
while $r_i$ is the distance between the i-th helium atom and the 
hydride ion.
In this work, for $V_{HeHe}(r_{ij})$ we employ the recent
Tang-Toennies-Yiu (TTY) pair potential ~\cite{tty} that is not based on any kind
of empirical information. This choice allows us to directly compare
our results
with the recent ones obtained by Lewerenz ~\cite{lewhe}, who used this pair
potential to compute energetics and structure of various small
$^4$He$_N$ clusters.
To obtain the pair potential between He and H$^-$ we fitted the accurate
Full CI results by Bendazzoli, Evangelisti and Passarini  ~\cite{bend} 
employing an analytical form
tailored to approximate with similar accuracy all the three regions of
the potential energy curve:

\begin{equation}
\label{cluseq3.1}
V_{HeH^-}(r) = a_1 r^{a_2} e^{-a_3r} + a_4   \;\;\;\;\;\;\;r <10 \;bohr
\end{equation}
\begin{equation}
\label{cluseq3.2}
V_{HeH^-}(r) = a_5 (1-e^{a_6(r-a_7)})^2 -a_8  \;\;\;\;\;\;\;10 \;bohr \leq r 
\leq 20 \;bohr 
\end{equation}
\begin{equation}
\label{cluseq3.3}
V_{HeH^-}(r) = \frac{a_9}{r^4}   \;\;\;\;\;\;\;\; r > 20 \;bohr 
\end{equation}
\noindent
This approach differs from the one used by Bendazzoli {\it et al.} in their
work, since they chose to interpolate their data using exponential splines
instead of fitting them with any analytical model.
To fit the 19 computed values of the interaction potential
with our analytical model,
we used the Levenberg-Marquard algorithm, imposing analytically 
the continuity  between the functions at 10 and 20 bohr. 
The parameters obtained by means of this procedure are (in atomic units):
$a_1 =1.303648$, $a_2 = -1.297418$, $a_3 = -0.7503146$, $a_4 = 0.00001989$,
$a_5 = 0.000012769$, $a_6 = 0.313155$, $a_7 = 13.0$, $a_8 = -0.00001467$
and $a_9 = 0.736841$.
Figure 1 shows both the fitted $V_{HeH^-}(r)$ and 
the TTY $V_{HeHe}(r)$ potentials.
From Fig. 1, one can note that the two potentials have quite different
well depth and location of the minimum: these results can be explained
remembering that H$^-$ is very diffuse, and that for this reason
the attractive charge-induced dipole interaction can occur only for large
internuclear separation. Therefore, the well depth should be in principle
quite small.
In this work we do not introduce any kind of information about
three-body forces: while for pure helium clusters one can use
the standard Axilrod-Teller term ~\cite{axi}
to augment the pair potential approximation,
to our knowledge similar information are not available for the
$^4$He$_2$H$^-$ trimer.

To approximate the ground state wave functions for these systems,
we employed the commonly used pair product form ~\cite{rick}

\begin{equation}
\label{cluseq3bis}
\Psi _T (\mathbf{R}) = \prod _{i<j}^N \psi (r_{ij})
\prod _i ^N \phi (r_{i})
\end{equation}
\noindent
and no one-body part was used. This fact guarantees that we do not
introduce any center-of-mass kinetic energy component in the description
of the cluster, avoiding us the burden of subtracting it to obtain
the internal energy of the system.
Both the $\psi (r)$ and $\phi (r)$ functions have the same analytical form

\begin{equation}
\label{cluseq4}
\psi (r) = \phi (r) = exp [ -\frac{p_5}{r^5} -\frac{p_3}{r^3}
-\frac{p_2}{r^2} -p_1 r -p_0 ln(r) ]
\end{equation}
\noindent
This form is identical to the one employed by Rick, Lynch and Doll
~\cite{rick} except for
the presence of the new term $-p_3/r^3$ that was previously used by
Barnett and Whaley ~\cite{whcluster} in their study of helium clusters. 
During the preliminary stages of our work, this term was found
to improve sensibly the variational energy of the wave function,
and to have a positive impact on the stability of the optimization.

The chosen form for the trial wave function makes  impossible to
compute analytically the matrix elements of the Hamiltonian operator,
and numerical methods must be used to obtain the
energy and other mean values for a given trial wave function. The
variational Monte Carlo method is well suited for this goal since it
requires only the evaluation of the trial wave function, its gradient,
and its Laplacian. Since this and other Monte Carlo methods are well
described in the literature ~\cite{reybook},  
we refer the reader to it and to our previous
work in this field for the details.
However, it is relevant to point out that  all the mean values were computed
by means of the general integral

\begin{equation}
\label{cluseq5}
\langle {\mathcal O} \rangle = \frac{\int f ({\mathbf R})
{\mathcal O}_{loc} ({\mathbf R}) d{\mathbf R}}
{\int f ({\mathbf R}) d{\mathbf R}}
\end{equation}

\noindent
where

\begin{equation}
\label{cluseq6}
{\mathcal O}_{loc} ({\mathbf R}) = \frac{{\mathcal O} \Psi_{T}({\mathbf R})}
{\Psi_{T}({\mathbf R})}
\end{equation}

\noindent
and $f ({\mathbf R}) = \Psi_{T} ^2 ({\mathbf R})$ for VMC, while
$f ({\mathbf R}) = \Psi_{T} ({\mathbf R}) \Psi_{0} ({\mathbf R})$ for DMC.
In VMC calculations, Eq. \ref{cluseq5} gives exactly the expectation
values of the ${\mathcal O}$ operator over the trial wave function $\Psi_{T}$, 
while in DMC simulations Eq. \ref{cluseq5} gives only an improved estimate
of the value, but not the exact one. This is true for all the ${\mathcal O}$ 
operators that do not commute with the Hamiltonian. 

As to our general strategy to optimize the trial wave functions for the 
clusters studied, we usually select the parameters 
of the exponential part
in Eq. \ref{cluseq4} by minimizing the estimate of the variance of the local
energy over a fixed sample of walkers

\begin{equation}
\label{cluseq7}
\sigma ^2 = \frac{1}{N_{walker}} \sum _{i=1}^{N_{walker}}
[E_{loc}(\mathbf{R}_i) -E_{ref}]^2
\end{equation}

All the VMC optimizations, and the VMC and DMC simulations were carried out
using at least 5000 walkers; all the DMC simulations were run employing
a time step of 200 hartree$^{-1}$ and the accuracy of the results
was checked running few more simulations with smaller time steps to ensure
that the time step bias was negligible for all the expectation values.

The wave functions are available from the authors upon request.

\section{RESULTS AND DISCUSSION}

In order to check our code,
we carried out DMC simulations on the small systems
$^4$He$_2$, $^4$He$_3$ and $^4$He$_4$ employing the parameter listed in 
Ref. ~\cite{rick} for the trial wave functions.
Our DMC results,
-0.00089(1) cm$^{-1}$ for $^4$He$_2$, -0.08784(7) cm$^{-1}$ for $^4$He$_3$ 
and -0.3886(1) cm$^{-1}$ for $^4$He$_4$, are in optimal agreement
with the results obtained by Lewerenz in Ref. ~\cite{lewhe}.
As far as $^4$HeH$^-$ is concerned, we optimized a wave function of
the form of Eq. \ref{cluseq4} obtaining  -0.36987(8) cm$^{-1}$ as mean energy.
Using this trial wave function we obtained a
DMC energy of -0.3969(4) cm$^{-1}$, that appears to be
in good agreement with the result of -0.4000 cm$^{-1}$
obtained by Bendazzoli {\it et al.} ~\cite{bend}
for the J=0 case.
The remaining small discrepancy is due to the different method we used to
obtain an analytical representation of the interaction potential, and show
the accuracy of the fitted potential.

These results allow a first comparison between $^4$He$_2$ and $^4$HeH$^-$:
although the well depth of the interaction potential energy 
between two helium atoms
is almost twice the well depth of the HeH$^-$ potential,
and the reduced mass of $^4$HeH$^-$ is smaller than the one of $^4$He$_2$,
the total energies differ by  more than two orders of magnitude favoring  the
stability of $^4$HeH$^-$. This outcome can be explained noticing that
the HeH$^-$ potential has a longer asymptotic decay and a shallower
well than the He$_2$ potential. These features reflect themselves
in a narrower wave function and a smaller mean distance between $^4$He and H$^-$
than between $^4$He and $^4$He.

Having tested our code,
we optimized a trial wave function for the smallest cluster
$^4$He$_2$H$^-$ starting from the parameters of the wave functions
of the two dimers $^4$He$_2$ and $^4$HeH$^-$. Since this initial wave function
was a crude approximation to the ground state, instead of using a VMC 
distribution of walkers to carry out the parameter optimization, 
we employed a DMC simulation to select the configurations. 
This alternative way, although seldom used,
has the advantage to push the distribution towards the correct one,
biasing the selection of the parameters
of the trial wave function toward better ones.
After a couple of optimization steps using the DMC
distributions, the wave function parameters
had roughly converged, allowing us to use VMC simulations to compute
mean values and to select the new configurations to carry out
the optimization procedure itself for this small system.

For all the $^4$He$_N$H$^-$ systems with $N>2$ the wave function
optimization was started using the parameters of the cluster having
one helium atom less. We found this choice to be a good initial guess
for the minimization procedure and a good distribution
to select the set of configurations by means of a VMC simulation.
The VMC results obtained by means of the optimization of the trial wave function
for the $^4$He$_N$H$^-$ are shown in Table I \ref{tab1}. 

Since our trial wave functions are only an approximation of the true ground
state functions and allow to compute only approximate properties of 
these clusters, to project out all the
remaining excited state contributions we employed 
DMC simulations to sample the distribution $f(\mathbf{R}) = \Psi _T
(\mathbf{R}) \Psi _0 (\mathbf{R})$.
The DMC energy and potential mean values are shown in Table I \ref{tab1}
together with the VMC results.
Comparing DMC and VMC energy results, one can note that the 
percentage of total energy  recovered by the VMC wave functions decreases
in monotonic fashion, starting from 96\% for the N$=2$ cluster
and ending to 81\% for the largest cluster N=12 studied in this work.
At present, we are not able to include any definitive explanation
of this behaviour, but we feel that it could be due either to the limitation
of the model function itself or to the optimization procedure based on
the variance of the local energy, or both.

Differently from what has been noted in Ref. ~\cite{lewhe} for the pure helium
clusters, energies and other mean values for the $^4$He$_N$H$^-$
systems converged quite easily even for the smallest clusters
with $N \leq 4$.

From the results shown in Table I \ref{tab1}, supplemented with the
DMC results by Lewerenz on pure helium clusters, one can compute
various interesting energetic quantities:

\begin{eqnarray}
\label{cluseq8}
E_{ex}(N) = E_{^4He_N} - E_{^4He_{N-1} H^-} \\ \nonumber
E_{grow}^{H^-}(N) = E_{^4He_{N-1} H^-} - E_{^4He_N H^-}  \\ \nonumber
E_{grow}^{He}(N) = E_{^4He_{N-1}} - E_{^4He_N}  \\ \nonumber
E_{bind}(N) = E_{^4He_N} - E_{^4He_N H^-}
\end{eqnarray}

\noindent
where $E_{ex}(N)$ represents the energy that is released exchanging an helium 
atom with the hydride ion, $E_{grow}^{H^-}(N)$ and $E_{grow}^{He}(N)$
the energes that are released adding
an $^4$He to an already formed $^4$He$_{N-1}$H$^-$
cluster or $^4$He$_{N-1}$ cluster respectively, 
while $E_{bind}(N)$ is the binding energy
of H$^-$ to the $^4$He$_N$ cluster.
These quantities are shown in Figures 2 and 3 to allow a quick comparison,
together with the total energy and $E_{grow}(N)$ for the pure helium
clusters obtained by Lewerenz. We supplemented his results with the 
total energy for $^4$He$_{11}$ (-7.288(3) cm$^{-1}$), $^4$He$_{12}$ (-8.746(7) 
cm$^{-1}$), $^4$He$_{13}$ (-10.299(4) cm$^{-1}$) computed in this study.

Similarly to the results obtained by Barnett and Whaley 
~\cite{whh2} in their work on
helium clusters containing an hydrogen molecule as impurity, from Figure 2 and 3
it is possible to note that the energetics of these small $^4$He$_N$H$^-$
clusters is dominated by the presence of the H$^-$ ion.
The total energy of the $^4$He$_N$H$^-$ appears to be
much lower than the energy of $^4$He$_N$.
For the clusters we studied, both $E_{bind}(N)$ and $E_{grow}(N)$ increase 
almost linearly with the number of helium atoms.
This is an expected result for $E_{grow}(N)$ since, if no three-body contribution
to the potential energy is present, the total energy of a cluster
should be roughly proportional to the number of pairs present.
As far as $E_{bind}(N)$ is concerned, its almost linear behavior cannot be
explained by means of a similar reasoning;
this outcome could be easily rationalised
if H$^-$ were solvated by the He atoms. Unfortunatly this appears
to be hardly possible, due to the quite different well minimum location
of the two potentials.

To obtain information about the structure of doped clusters, 
during the DMC simulations we collected the radial distribution $R(r)$
from the center of mass for both $^4$He and H$^-$ 

\begin{equation}
\label{cluseq9}
\mathbf{R}_{CM} = \frac{m_{^4He} \sum _{i=1} ^{N} \mathbf{r}_i + 
m_{H^-} \mathbf{r}_{H^-}}{N_{He}m_{^4He} + m_{H^-}}
\end{equation}
\noindent
and from the geometrical center of the cluster
\begin{equation}
\label{cluseq10}
\mathbf{R}_{G} = \frac{\sum _{i=1} ^{N} \mathbf{r}_i + 
\mathbf{r}_{H^-}}{N_{He} + 1}
\end{equation}

Figure 4  and 5 display the results for  $R(r)$
respect to the geometrical center; these are normalized such that

\begin{equation}
\label{cluseq11}
\int _0 ^{\infty} R(r)_{He} r^2 dr = N_{He}
\end{equation}
\noindent
for the helium atoms, while

\begin{equation}
\label{cluseq12}
\int _0 ^{\infty} R(r)_{H^-} r^2 dr = 1
\end{equation}
\noindent
for the H$^-$ ion.
We chose to show only $R(r)$ with respect to the geometrical center,
since the same quantity computed with respect to the center of mass does not
introduce any new information.

Comparing the He radial density distributions shown in Figure 4 with the same
profiles obtained by Lewerenz ~\cite{lewhe}, 
it is possible to note that they appear quite
similar except for our three-body cluster, i.e. $^4$He$_2$H$^-$.
The He density distribution for this cluster shows a maximum around 5.30 bohr,
but there is no trace of the rise of the density
for small distances from the center that can be seen in the 
case of $^4$He$_3$.
Nevertheless, the plot of Figure 4 shows that He can occupy the geometrical
center position, i.e. $^4$He$_2$H$^-$ in its ground state can be found in
the linear geometry where the H$^-$ ion is external to the $^4$He$_2$ moiety.
Increasing the number of He atoms present in the cluster, the density near
the cluster center rises toward the bulk value, represented in Figure 4
by the horizontal dotted line. 
Similarly to the pure He clusters, the helium atoms
appear to be completely delocalized with no indication of any
shell filling structure.

From the H$^-$ density distributions shown in Figure 5, one can note the peaked
distribution of the impurity with respect to the shallower profile of the He 
atoms, and that, upon
increasing the number of helium atoms in the cluster, the H$^-$
is pushed toward larger distances from the center. Moreover, 
Figure 5 shows that
the penetration of H$^-$ decreases in a fairly monotonic fashion.
The only exception to this behavior is the smallest cluster
$^4$He$_2$H$^-$: its distribution shows
a rise beyond statistical fluctuation for a distance from the center
less than the minimum located around 4.1 bohr. This result indicates that in
the $^4$He$_2$H$^-$ cluster the linear geometry where H$^-$
lies between the He atoms plays a significant role in the description
of the total motion, although H$^-$ has a larger probability to lie 
10 bohr far from the center of the cluster. 
Concluding, except for the smallest
cluster of the series, the impurity is not solvated by the $^4$He atoms, having
a small probability to be found near the center of the cluster. 
Analogous conclusions
can be obtained by the density distributions computed respect to the center of 
mass of the cluster.

The pair distribution functions $P(r)$ of the $^4$He-$^4$He and $^4$He-H$^-$ distances
are shown in Figures 6 and 7, respectively. These distributions are normalized
such that $\int _0 ^{\infty} p(r) r^2 dr = N_{He}$.
Again, the He-He distribution has many similarities with the same quantity
computed in Ref. ~\cite{lewhe}, showing a short range structure given by the 
sharp boundary hole around each He atom and a long decaying tail for
large distances. Moreover, for N$\geq 10$, our pair distributions show
a weak shoulder around 12-13 bohr; the onset of this peculiarity of the
He-He pair distribution was already noted by Barnett 
and Whaley ~\cite{whcluster} for $^4$He$_{13}$ and $^4$He$_{14}$ in their
work and explained by means of the appearence of the 
second-nearest-neighbor coordination shell.
This cannot be the case for our N=10-12 clusters,
where the icosahedral  shell filling is not even completed.
A similar feature seems to appear for $^4$He$_{10}$ in the
pair distribution shown by Lewerenz in his work on pure clusters ~\cite{lewhe}.
We interpret this feature as due to the presence of a light residual of 
an almost icosahedral structure of the clusters, i.e. an $^4$He
does not have only first neighbors.
Moreover, we suspect this trait to be emphasized by the 
light deformation of the $^4$He$_N$
moiety distribution due to the impurity resident on its surface.
The presence of this deformation for all the studied clusters
is supported by the almost linear behavior of E$_{bind}$ shown in Figure 2,
and by the fact that only small changes take place in the 
form of the He-H$^-$ distributions (Figure 7)
upon increasing of the number of atoms in the clusters.

From the sampled density and pair distributions, displayed in Figures 4-7,
various mean distances can be computed by simple one dimensional integration.
In Table II \ref{tab2}, 
we report the values for $<r_{HeH^-}>$, $<r_{HeHe}>$, $<r_{HeCM}>$,
and $<r_{H^-CM}>$. Comparing our $<r_{HeHe}>$ and $<r_{HeCM}>$ results
with the mean values computed by Lewerenz for the clusters containing
the same total number of particles, we note that our doped clusters have
a more compact structure of the $^4$He atom moiety than the pure ones,
certainly due to the presence of the strongly binding impurity.
Both mean distances show a sharp decrease up to N=5  where a minimum is located,
and a light increase going toward larger N. This behavior can be explained
easily by means of standard arguments: 
going from N=2 to N=5 the mean interaction
between the particles increases due to their increased number, giving rise
to a stronger binding between them. For N greater than 5 the cluster becomes
larger due to the addition of another $^4$He atom, in spite of the augmented
total interaction energy. A similar behavior is displayed by the $<r_{HeH^-}>$
mean values, and it can be rationalized by means of the same arguments.

As far as $<r_{H^-CM}>$ is concerned, its monotonic increase 
going toward larger N
can be explained by the joined effect of the increased dimension of the 
$^4$He atoms moiety and of the displacement of the center of mass location
inside the moiety itself, due to a simple mass effect.

\section{CONCLUSIONS}

In this work we presented the first quantum-mechanical study of an anionic
impurity in $^4$He clusters. Our results show the H$^-$ ion to
be located on the surface of the cluster, except for the $^4$He$_2$H$^-$
cluster where it has a finite probability to be found between the two
He atoms in a linear geometry. Our total energy values show that
the interaction between helium and the impurity is an important
component of the total energy of the system, while our mean geometrical
values show that the helium moiety is slightly contracted with respect to the
pure cluster case. We consider this fact to be due to the
form of the interaction potential between an helium atom and the impurity,
and especially to its longer tail respect to the He-He potential.

As far as the possibility to record a microwave spectrum of these systems is
concerned, the location of the impurity on the cluster surface
seems to indicate that this is possible, at least in principle.
In fact, the cluster structures resemble mostly the structure of a 
heteronuclear diatomic molecule whose lightest atom carries a negative charge,
and whose heaviest atom has a farly large radius.

In this work, we approximated the total interaction potential as a sum
of pair components, therefore excluding any non-addictive effect in the 
description of the systems.
The three-body interaction between two He and one H$^-$ 
could play a major role than in the pure He case
in defining both energetics and
structure of the clusters, due to the fact that it is due to a
charge-induced dipole-induced dipole interaction.

Moreover, the effects of the isotopic substitution of H$^-$ with
D$^-$ on the structure and energetics are worth to study,
since they were found quite important for
HeH$^-$ ~\cite{bend}. We expect this effect to be especially
important for the excited rotational states of these complexes,
so we are planning to extend our study in this new direction,
i.e. to compute ground state properties for clusters containing
D$^-$, and rotational excited states for both the doping
impurities ~\cite{whrot}.

\section*{ACKNOWLEDGMENTS}

This work was supported by the italian MURST grant N. 9803246003.
The authors wish to thank G. L. Bendazzoli for providing the
interaction potential values between He and H$^-$.
The authors are also indebted to the
Centro CNR per lo Studio delle Relazioni
tra Struttura e Reattivita' Chimica for grants of computer time.
Also, this work has benefited from a Post-Doctoral fellowship of MM.

\pagebreak

\begin{table}
\begin{center}

\begin{tabular}{lcccccc}  \hline\hline
N  & E$^{VMC}_0$ & V$^{VMC}_0$ & $\sigma$  & E$^{DMC}_0$ & V$^{DMC}_0$&E$^{VMC}_0/$E$^{DMC}_0$\\ \hline
2  & -1.0565(15) & -2.9628(87) & 1.291(11) & -1.0912(7)  & -2.9804(21)&0.97\\
3  & -1.9673(2)  & -5.7457(22) & 1.927(17) & -2.0476(20) & -5.7918(65)&0.95\\
4  & -3.0392(15) & -9.633(11)  & 2.458(22) & -3.2569(66) & -9.656(22) &0.94\\
5  & -4.3718(21) & -12.424(13) & 2.941(65) & -4.6725(43) & -13.280(13)&0.93\\
6  & -5.8072(43) & -17.601(15) & 3.928(44) & -6.2570(44) & -18.628(19)&0.93\\
7  & -7.3083(21) & -22.314(13) & 4.433(43) & -7.9667(65) & -23.219(22)&0.91\\
8  & -8.8820(44) & -28.356(20) & 5.421(44) & -9.799(18)  & -29.211(66)&0.90\\
9  &-10.5718(65) & -35.956(17) & 7.023(65) & -11.763(22) & -35.70(13) &0.89\\
10 &-12.2025(66) & -39.329(22) & 7.638(21) & -13.850(11) & -40.73(12) &0.88\\
11 &-13.703(11)  & -48.305(44) & 9.437(64) & -15.990(17) & -48.54(16) &0.85\\
12 &-14.900(17)  & -58.27(11)  & 11.588(65)& -18.220(15) & -57.21(22) &0.82\\

\hline \hline

\end{tabular}
\caption{VMC and DMC energy and mean potential 
results for the $^4$He$_N$H$^-$ clusters.
All energy values are in cm$^{-1}$.
}
\label{tab1}
\end{center}
\end{table}

\newpage

\begin{table}
\begin{center}

\begin{tabular}{lcccc}  \hline\hline
N$_{He}$ &  $<r_{HeH^-}>$& $<r_{HeHe}>$& $<r_{H^-CM}>$ & $<r_{HeCM}>$\\ \hline

2   & 21.004  & 14.856   & 12.997   & 9.923 \\
3   & 20.380  & 13.234   & 14.153   & 8.941 \\
4   & 20.042  & 12.224   & 14.942   & 8.279 \\
5   & 19.964  & 12.210   & 15.251   & 8.263 \\
6   & 20.002  & 12.459   & 15.811   & 8.392 \\
7   & 20.172  & 12.621   & 16.221   & 8.567 \\
8   & 20.019  & 12.590   & 16.333   & 8.570 \\
9   & 20.257  & 12.798   & 16.717   & 8.674 \\
10  & 20.574  & 12.983   & 17.123   & 8.874 \\
11  & 20.547  & 12.920   & 17.236   & 8.889 \\
12  & 20.513  & 12.936   & 17.332   & 8.883 \\ \hline \hline

\end{tabular}
\caption{DMC mean values for observables
of the $^4$He$_N$H$^-$ clusters. All values are in bohr.}
\label{tab2}
\end{center}
\end{table}

\newpage

\newpage
\section*{Figure captions}

\noindent
Figure 1: He-He and He-H$^-$ interaction potentials ($\mu$hartree).

\noindent
Figure 2: $E_{bind}$, $E_{grow}^{H^-}$, and $E_{ex}$ vs. the number of He atoms
for the $^4$He$_N$H$^-$ clusters. Energies are in cm$^{-1}$.

\noindent
Figure 3: Total energy and $E_{grow}$ for both pure He and H$^-$ doped clusters.
Energies are in cm$^{-1}$.

\noindent
Figure 4: Radial density distribution of He atoms respect 
to the geometrical center of the clusters.

\noindent
Figure 5: Radial density distribution of H$^-$ ion respect 
to the geometrical center of the clusters.

\noindent
Figure 6: He-He pair distribution in the clusters.

\noindent
Figure 7: He-H$^-$ pair distribution in the clusters.

\pagebreak

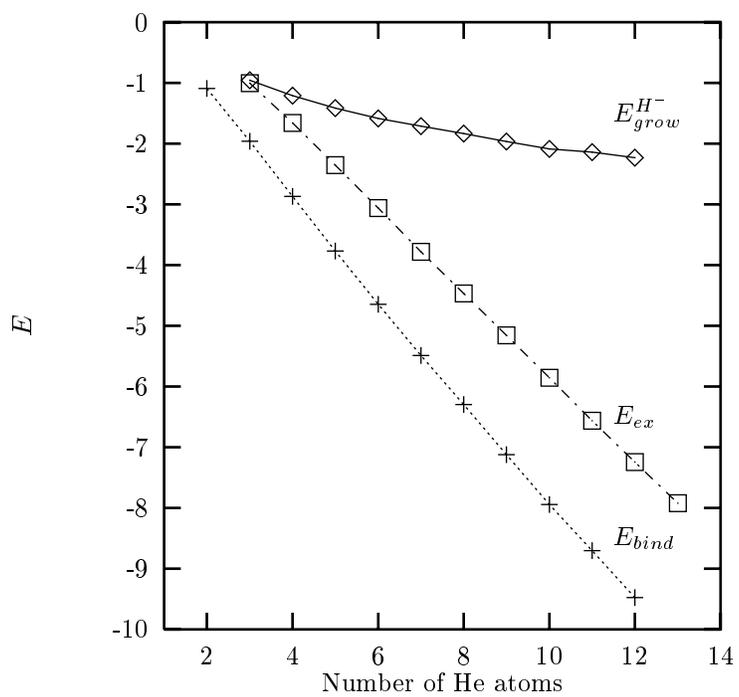
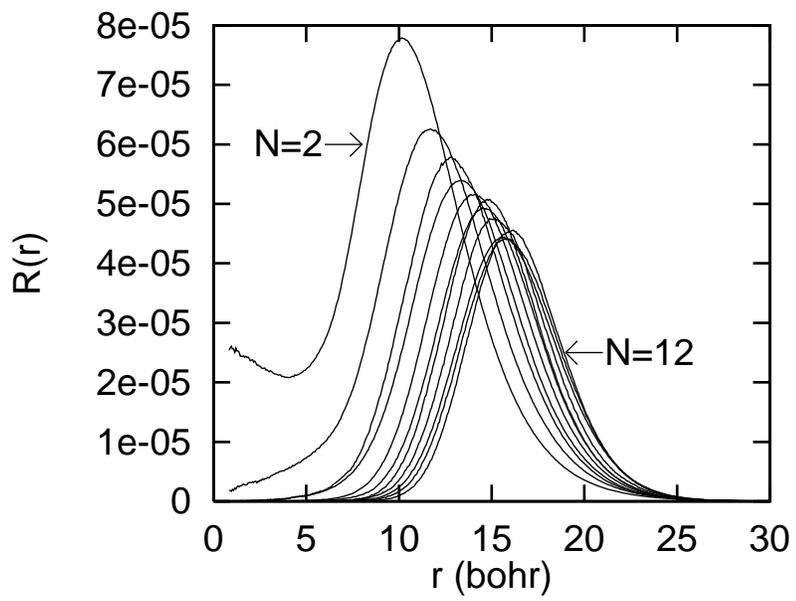
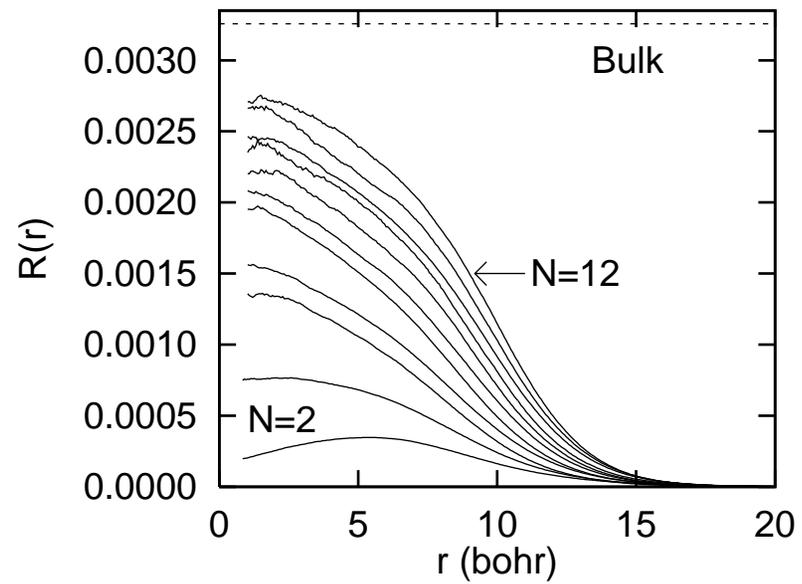
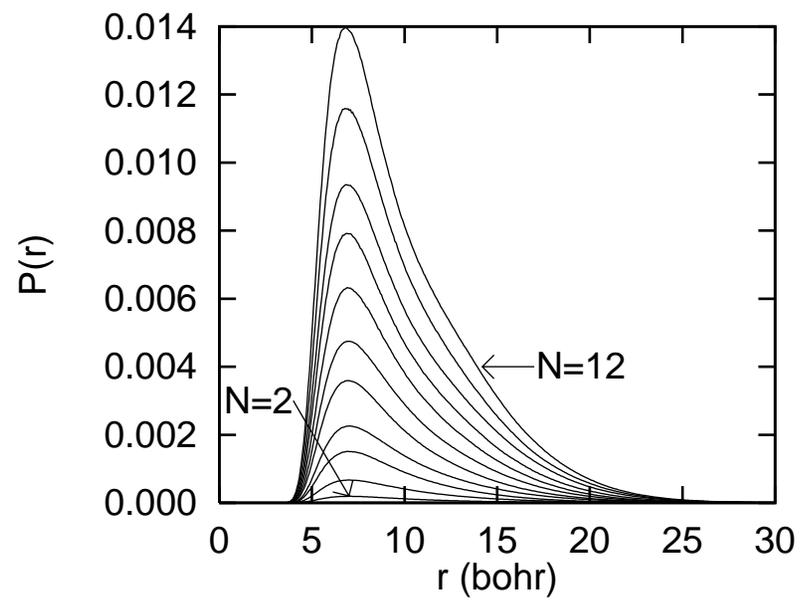
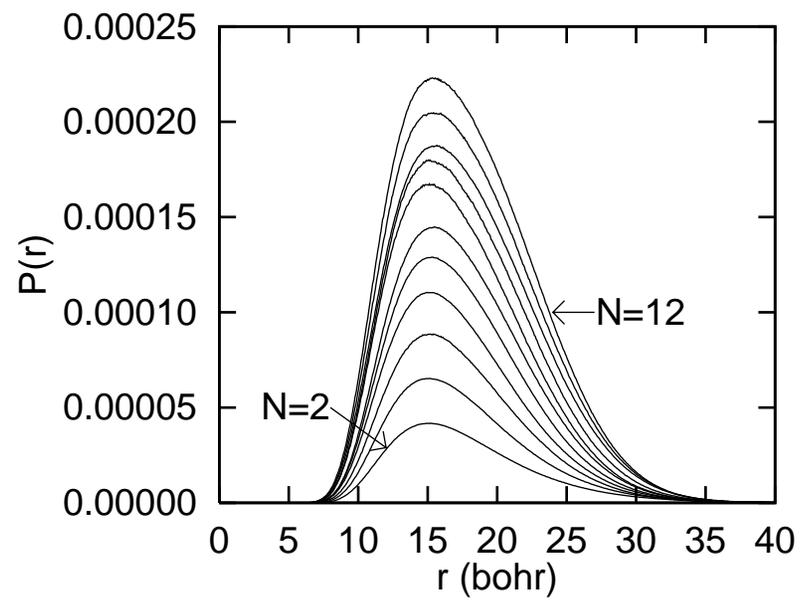
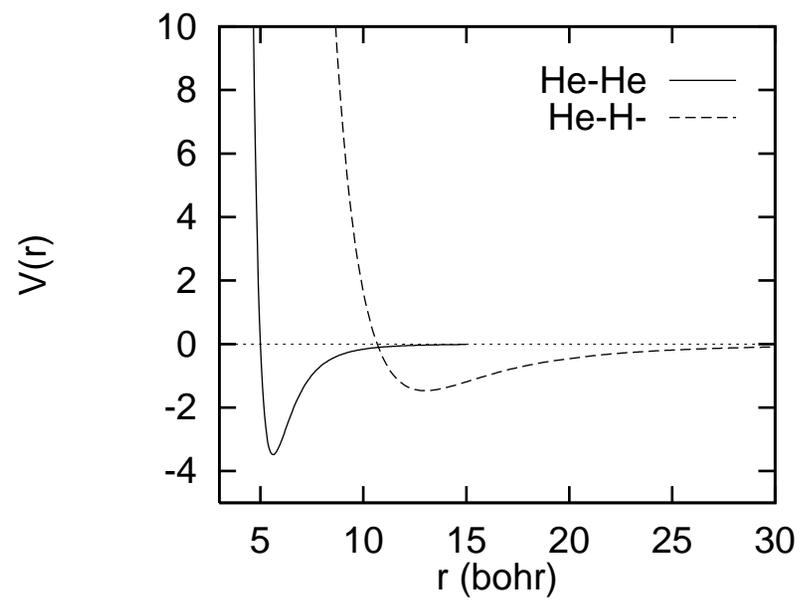
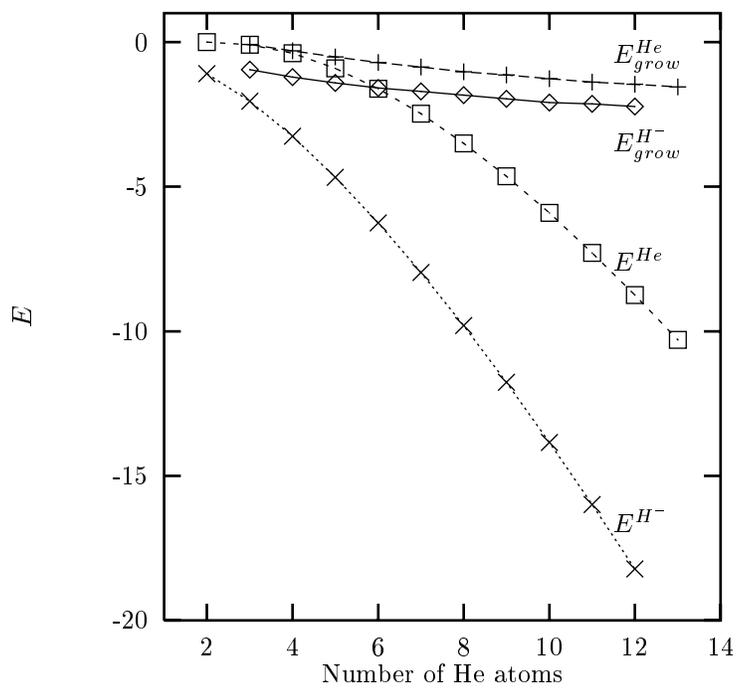
\end{document}